\documentclass[letterpaper]{article} 
\usepackage{aaai2026}
\usepackage{times}  
\usepackage{helvet}  
\usepackage{courier}  
\usepackage[hyphens]{url}  
\usepackage{graphicx} 
\usepackage{natbib}  
\usepackage{caption} 
\usepackage{algorithm}
\usepackage{algorithmic}
\usepackage{booktabs}
\usepackage{tcolorbox}
\usepackage{amsmath}
\usepackage{amsfonts}
\usepackage{newfloat}
\usepackage{listings}
\DeclareCaptionStyle{ruled}{labelfont=normalfont,labelsep=colon,strut=off} 
\floatstyle{ruled}
\newfloat{listing}{tb}{lst}{}
\floatname{listing}{Listing}
\title{A Closed-Loop Multi-Agent Framework for\\Aerodynamics-Aware Automotive Styling Design}

\author{
    Xinyu Jin\textsuperscript{\rm 1},
    Shengmao Yan\textsuperscript{\rm 1},
    Qingtao Wang\textsuperscript{\rm 1},
    Shisong Deng\textsuperscript{\rm 1}, \\
    Yanzhen Jiang\textsuperscript{\rm 1},
    Shuangyao Zhao\textsuperscript{\rm 1,2}\thanks{Corresponding author.}
}

\affiliations{
    \textsuperscript{\rm 1}School of Management, Hefei University of Technology, Hefei, China\\
    \textsuperscript{\rm 2}Key Laboratory of Process Optimization and Intelligent Decision Making, Hefei, China\\
}

\usepackage{bibentry}

\begin{document}

\maketitle

\begin{abstract}
The core challenge in automotive exterior design is balancing subjective aesthetics with objective aerodynamic performance while dramatically accelerating the development cycle. To address this, we propose a novel, LLM-driven multi-agent framework that automates the end-to-end workflow from ambiguous requirements to 3D concept model performance validation. The workflow is structured in two stages: conceptual generation and performance validation. In the first stage, agents collaborate to interpret fuzzy design requirements, generate concept sketches, and produce photorealistic renderings using diffusion models. In the second stage, the renderings are converted to 3D point clouds, where a Drag Prediction Agent, built upon a lightweight surrogate model, provides near-instantaneous predictions of the drag coefficient and pressure fields, replacing time-consuming CFD simulations. The primary contribution of this work is the seamless integration of creative generation with a rapid engineering validation loop within a unified, automated system, which provides a new paradigm for efficiently balancing creative exploration with engineering constraints in the earliest stages of design.
\end{abstract}

\section{Introduction}
\begin{figure}[t]
\centering
\includegraphics[width=0.9\columnwidth]{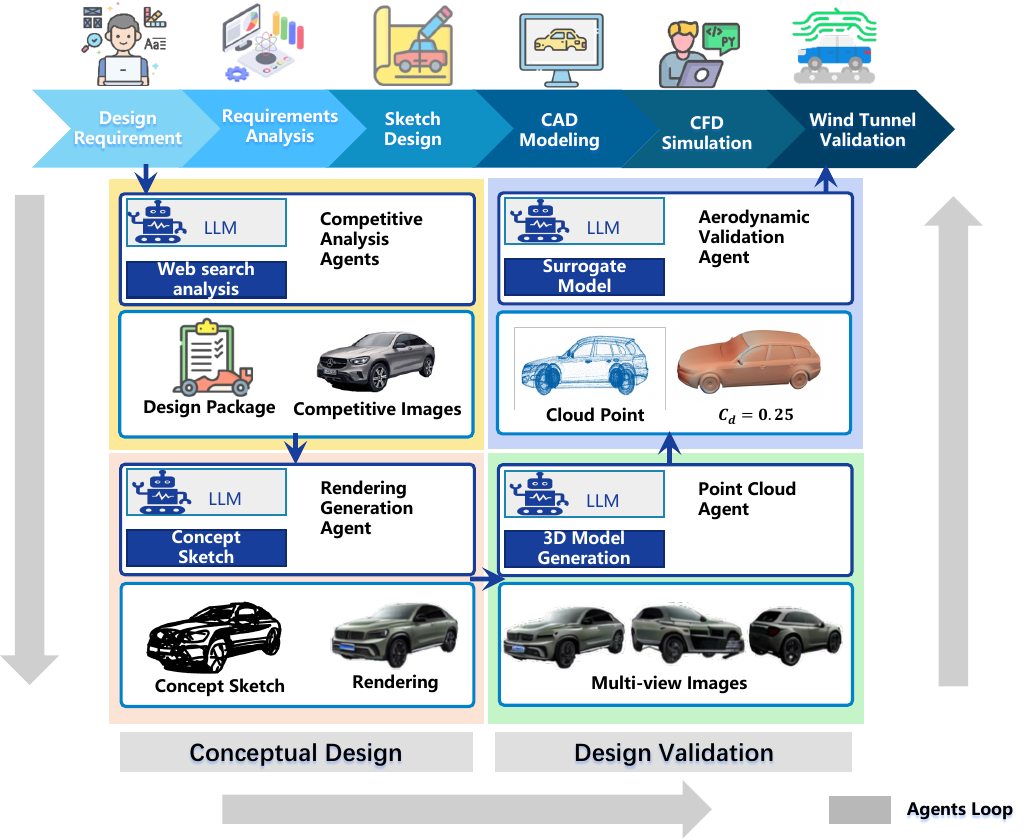}
\caption{Illustration of the proposed multi-agent framework. Our framework contributes to the early design exploration and rapid aerodynamic validation compared to traditional methods.}
\label{fig1}
\end{figure}

Automotive exterior design remains a complex negotiation between stylistic appeal and aerodynamic efficiency.  Visually striking forms often compromise drag and stability, while purely physics‑driven shapes may not meet market expectations.  As shown in Figure \ref{fig1} conventional workflows attempt to reconcile this tension through sequential phases: requirements analyses, designers sketch concepts, digital modellers create three‑dimensional representations, and engineers conduct computational fluid dynamics (CFD) simulations and wind‑tunnel tests.  Because CFD runs can take hours or days, rigorous aerodynamic validation typically occurs late in the process, forcing costly redesigns and compromises when poor performance is discovered.

Recent advances in generative models \citep{ddpm} and Large Language Model (LLM) driven multi‑agent systems \citep{autogen} offer an opportunity to rethink this pipeline.  Diffusion models yield photorealistic images conditioned on text or sketches, and LLM frameworks such as AutoGen enable specialized agents to collaborate on complex tasks.  However, existing design pipelines either apply generative models in isolation, leaving a gap to engineering validation, or rely on single‑agent systems whose scalability and robustness remain unclear.  In contrast, our work introduces a \emph{closed‑loop} design–validation workflow that seamlessly couples creative exploration with rapid aerodynamic assessment.  The key insight is to embed generative and surrogate models within an LLM‑orchestrated multi‑agent system, so that high‑level design prompts automatically trigger concept generation, 3D reconstruction and near‑real‑time aerodynamic prediction, which can dramatically accelerate the design process.

Compared with recent work \citep{mit}, our framework introduces several key improvements and innovations. First, during the CAD modeling stage, we reconstruct point cloud models from multi-view images instead of retrieving matches from existing models. This approach offers greater flexibility and practical value in real-world design workflows. Second, we adopt a fixed yet extensible agent-based workflow rather than relying on a LLM to coordinate all agent operations. This modular design enhances system stability, interpretability, and scalability, making it well-suited for evolving design pipelines. In addition, we incorporate a physics-attentive surrogate model capable of simultaneously predicting key aerodynamic indicators, including drag, pressure distribution, and velocity fields. The proposed model demonstrates superior performance on standard automotive aerodynamic benchmarks.

This paper introduces a novel framework that significantly advances the automotive design process by integrating LLM within a multi-agent collaborative system. Our contributions can be summarized as follows:
\begin{itemize}
    \item The seamless and continuous integration of "Aesthetic-to-Performance": Our system establishes direct bridge between subjective aesthetic attributes and quantifiable aerodynamic metrics. The multi-agent system can inherently incorporate various engineering constraints from the initial conceptualization, ensuring that aesthetically pleasing designs are also technically viable and performant.
    \item Specialized Agent Autonomy and Collaboration: Distinct agents (e.g., the "Competitive Analysis Agents") are empowered with specific domain findings and can autonomously execute tasks.
    \item Facilitate rapid prototyping and exploration: The generative capabilities of the multi-agent system, combined with its understanding of design principles and constraints, enable the rapid generation of diverse design variations that adhere to initial aesthetic goals while considering performance implications from the outset.
\end{itemize}

\section{Related Work}
\subsection{Generative Models in Concept Design}
In recent years, Deep Generative Models have undergone explosive development. Early research methods typically utilized VAEs \citep{vae} and GANs \citep{gan,stylegan} to generate novel images, while the recent wave of technologies represented by Diffusion Models \citep{ddpm} marks a paradigm shift. These new models have achieved unprecedented levels of quality, diversity, and controllability in generated content, advancing AI image generation from a phase of "random creation" to a practical stage of being "controllable, usable, and customizable" \citep{flux}. The latest advancements, exemplified by models like Stable Diffusion \citep{Stable} and DALL-E \citep{dalle}, fully demonstrate the immense potential of this field. Not only can they create detail-rich images based on complex textual instructions (Text-to-Image), but they have also fostered a vast community ecosystem. For instance, plugins like ControlNet \citep{contral} allow users to exert precise geometric control over the generation process through inputs such as sketches (Sketch-to-Image) or poses. In creative fields such as automotive design, researchers have leveraged these advanced models to rapidly generate a vast number of conceptual visuals, significantly accelerating the creative ideation phase and effectively exploring the aesthetic solution space \citep{burnap_car, vehiclesdf}. However, a notable limitation of such work is its isolated focus on visual generation. The outputs are typically static images, lacking a direct pathway to engineering analysis, which leaves the critical loop between aesthetics and performance disconnected.

\subsection{Surrogate Models for Aerodynamic Prediction}
To address the issues of traditional CFD methods being computationally expensive, time-consuming, and difficult to integrate into generative design processes, researchers are utilizing data-driven surrogate models to significantly increase simulation speed while maintaining high accuracy \citep{pointnet}. Early surrogate models predominantly relied on parameterized representations for performance prediction. For instance, vehicle profiles were characterized using discrete sets of control-point coordinates, and computational models were subsequently trained to estimate corresponding drag coefficients \citep{GUNPINAR201965, ROSSET2023142772}. Nevertheless, this method tends to oversimplify the geometric complexity of the automobile's shape, thereby limiting its applicability in practical design environments. To enhance shape feature extraction and representation, subsequent iterations of surrogate models have investigated learning-based methods utilizing two-dimensional or three-dimensional representations \citep{fengtian}. For example, three-dimensional automotive geometries are projected onto two-dimensional planes from various viewpoints, and implicit representations are learned using sophisticated neural network architectures, such as irregular convolutional neural networks or graph neural networks, to predict the drag coefficient \citep{song2023surrogatemodelingcardrag, Jacob_2022} and neural network-based surrogates can be deployed at industrial scale \citep{neuralcfd}. Additionally, generative models have been employed to reconstruct shapes from latent representations and predict pressure fields and drag coefficients, facilitating rapid and precise performance assessments \citep{Saha_2021, he2025drivaer}.
\begin{figure*}[t]
\centering
\includegraphics[width=0.9\textwidth]{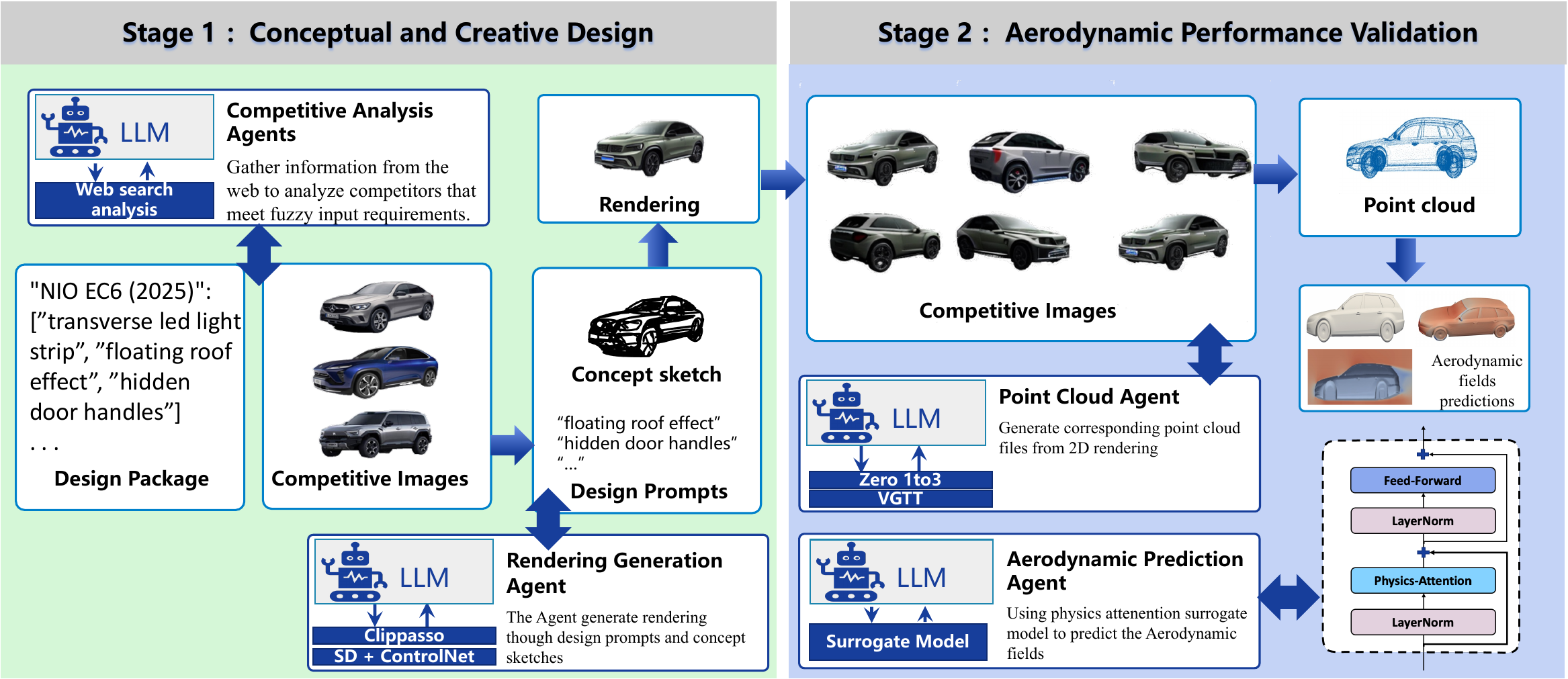} 
\caption{Overview of the proposed two-stage multi-agent framework. The first stage focuses on conceptual design, where user requirements are translated into visual concepts. The second stage emphasizes aerodynamic performance validation, utilizing surrogate models for rapid feedback.}
\label{fig:method}
\end{figure*}

\subsection{LLM-Powered Multi-Agent Systems}
Traditional design and simulation workflows often rely on rigid pipelines, siloed expert knowledge, and manual iterations, which severely limit scalability, adaptability, and cross-domain collaboration. Multi-Agent Systems (MAS) have emerged as a powerful paradigm for solving complex, distributed problems by enabling autonomous agents to collaborate, plan, and execute tasks \citep{multi-agent-co2}. With the rapid development of LLMs, the design and orchestration of such systems have entered a new era. LLMs now act as versatile cognitive engines that empower agents with advanced capabilities in reasoning \citep{react}, planning \citep{reflexion}, tool use \citep{toolformer}, and inter-agent communication \citep{LIU2025103643, multi-agent-corr}. This shift has led to the emergence of numerous LLM-centric multi-agent frameworks, such as AutoGen \citep{autogen}, LangChain \citep{ZHANG2025106244}, and ChatDev \citep{chatdev}, which enable the dynamic coordination of agents specialized in different tasks across domains including document processing \citep{proagent}, software development, and robotic control. In the design and simulation domain, recent efforts have explored the use of LLM-powered agents to automate multidisciplinary workflows. For instance, using agent assistants for semiconductor manufacturing \citep{Intelligent}. A recent work proposed a pioneering multi-agent system for joint aesthetic and aerodynamic car design \citep{mit}.

\section{Method}
\subsection{Overview: A Two-stage Approach}
In this section, we present a two-stage design process, illustrated in Figure \ref{fig:method}, built upon the AutoGen multi-agent framework, deconstructs the complex automotive design process into a logical, automated workflow. Stage one focuses on conceptual design, translating ambiguous user requirements into photorealistic renderings. Stage two is dedicated to rapid aerodynamic performance validation, transforming the rendering into an analyzable 3D model and using a lightweight surrogate model for near-instantaneous feedback. This creates a rapid, closed loop from creative concept to engineering validation.

\subsection{Conceptual and Creative Design}
The primary objective of this stage is to rapidly transform a designer’s abstract ideas into a diverse portfolio of high-fidelity, visually compelling concept designs. Traditionally, this process has depended heavily on manual efforts—designers would conduct extensive market research, draft preliminary sketches, and carry out detailed rendering, often requiring significant time and domain expertise. In contrast, our proposed multi-agent framework fully automates this pipeline by decomposing the task and distributing it among specialized agents. Specifically, the Competitive Analysis Agents performs real-time market analysis to identify trends, benchmarks, and differentiators, while the Rendering Generation Agent translates design intentions into high-quality visual outputs. 

\subsubsection{Competitive Analysis Agents} 
\begin{figure}[h]
\centering
\includegraphics[width=0.9\columnwidth]{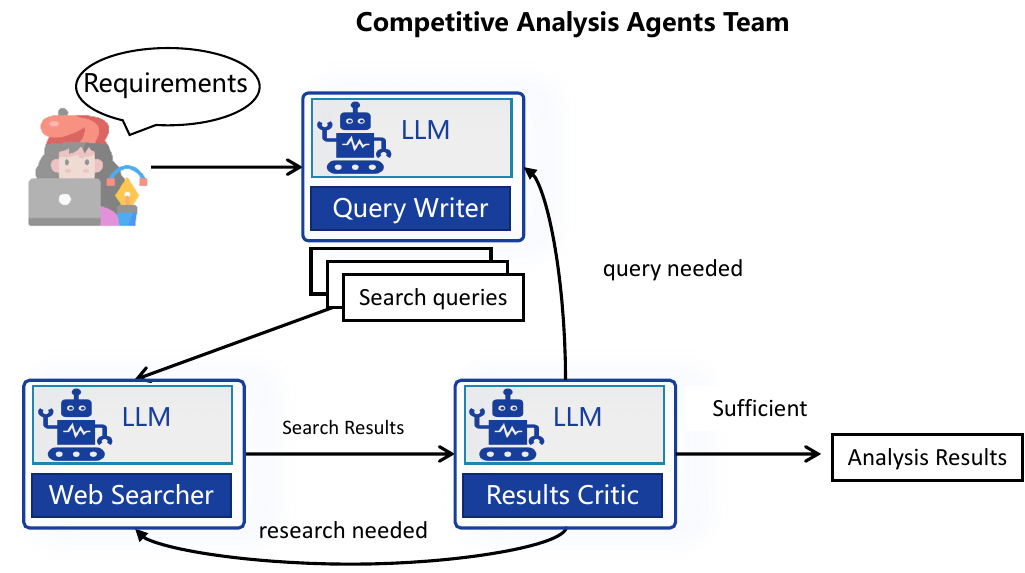}
\caption{search-reflect iterative loop of the Competitive Analysis Agent.}
\label{fig:search-reflect}
\end{figure}

The Competitive Analysis Agents comprises a team of collaborative sub-agents that translate high-level design requirements (e.g., a sporty SUV for the Chinese market with an aggressive appearance) into actionable market insights. Instead of performing a superficial, one-time search, it follows the "search-reflect" iterative loop shown in the Figure \ref{fig:search-reflect}, continuously optimizing queries based on ongoing analysis and deepening its dive into market information until a comprehensive insight is formed.
\begin{tcolorbox}[title={Generated Exploration Plan Example}]
\small
{
    \textbf{requirement}: ``A sporty SUV in the 300-500k RMB price range with an aggressive exterior, powerful performance, and a tech-focused interior'',\\
    \textbf{queries}: [``300-500k RMB sporty SUV recommendations 2024'', ``Top-rated SUVs with aggressive exteriors 300-500k RMB'', ``Powerful performance SUV model comparison 300-500k RMB'', ``Latest reviews of SUVs with high-tech interiors'']
}
\end{tcolorbox}

Upon completion of the loop, the agents synthesize their research into a structured design package, which systematically maps key competitors to the high-level design prompts derived from their distinctive features. This transforms raw market data into an organized and actionable format. The resulting output—comprising semantically rich design prompts and a curated image library for visual reference—forms a robust, data-driven foundation for the subsequent creative sketching phase. This ensures that the final design is both aligned with designer intent and attuned to prevailing market trends.
\begin{tcolorbox}[title={Design Package}]
\small
{
\texttt{"NIO EC6 (2025)": }["transverse led light strip", "floating roof effect", "hidden door handles"]\
\texttt{"Mercedes-Benz GLC Coupe (2025)": } ["luminous grille", "fastback silhouette"]\
\texttt{ "Mengshi M817 (2025)": } ["luminous grille", "hardcore off-road styling"]\
\dots
}
\end{tcolorbox}

\subsubsection{Rendering Generation Agent} 
After the "design package" is produced, the workflow is handed over to the Rendering Generation Agent. This agent's task is to creatively visualize the design concepts by generating high-fidelity renderings based on the insights provided by the Competitive Analysis Agents.

\begin{figure}[h]
\centering
\includegraphics[width=0.9\columnwidth]{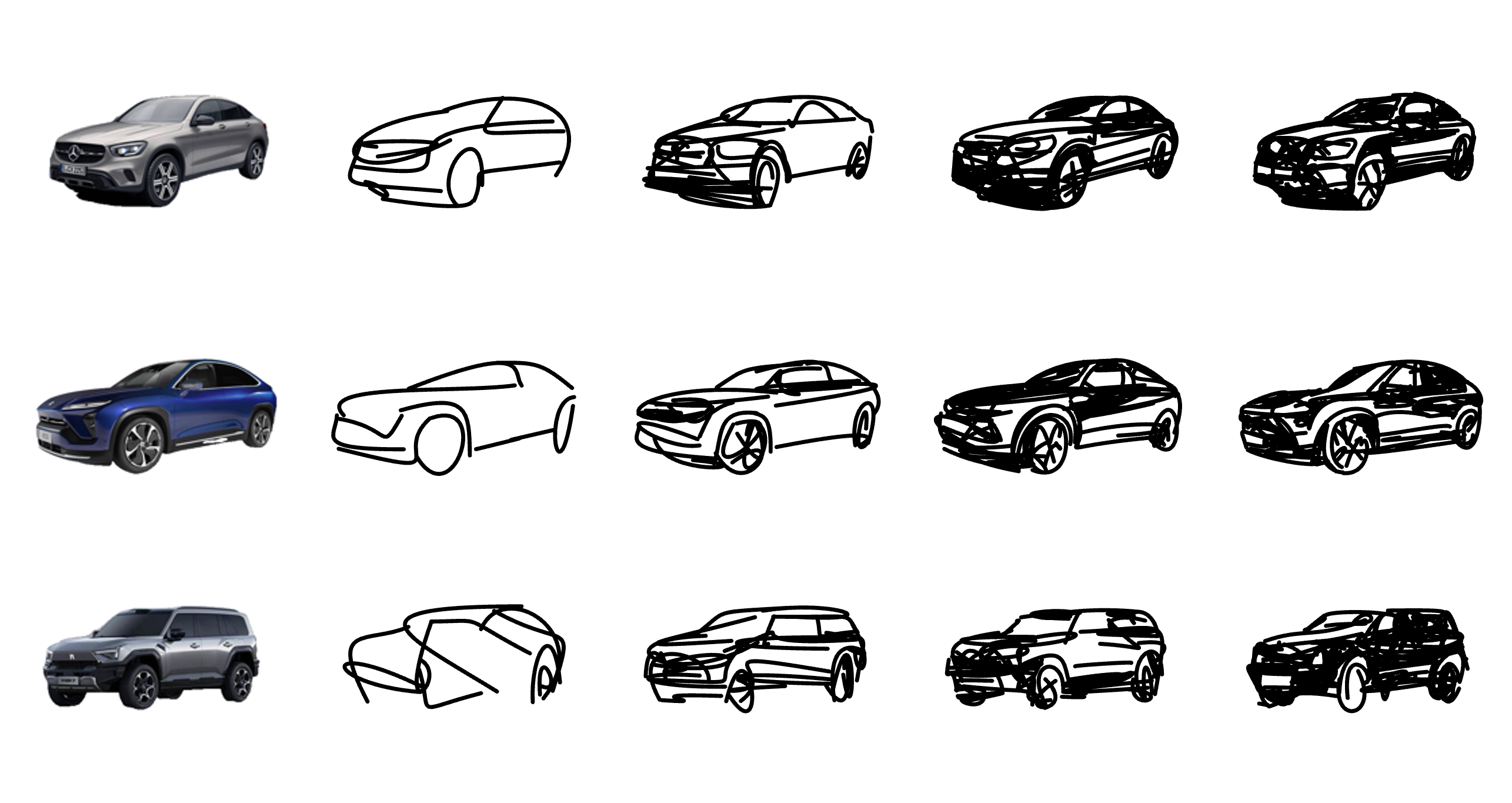} 
\caption{Concept sketches generated from a source image (left) via the Clipasso model, using 10, 25, 50, and 100 strokes to control the level of abstraction.}
\label{sketch}
\end{figure}

The first step for this agent is to generate a diverse set of concept sketches. It uses the Clipasso model to process the image library of competitor vehicles obtained in the previous step, generating a series of concept sketches in various artistic stroke styles, as shown in Figure \ref{sketch}, to provide designers with rich, unconventional early-stage creative inspiration.

Subsequently, the agent converts these sketches into photorealistic renderings. This process synergistically employs ControlNet and Stable Diffusion models to achieve a deep fusion of the Structural Information embodied in the concept sketch and the Semantic Information represented by the design prompts, resulting in a high-fidelity primary visual rendering.

\subsection{Aerodynamic Performance Validation}
\subsubsection{Point Cloud Agent}
\begin{figure}[h]
\centering
\includegraphics[width=0.9\columnwidth]{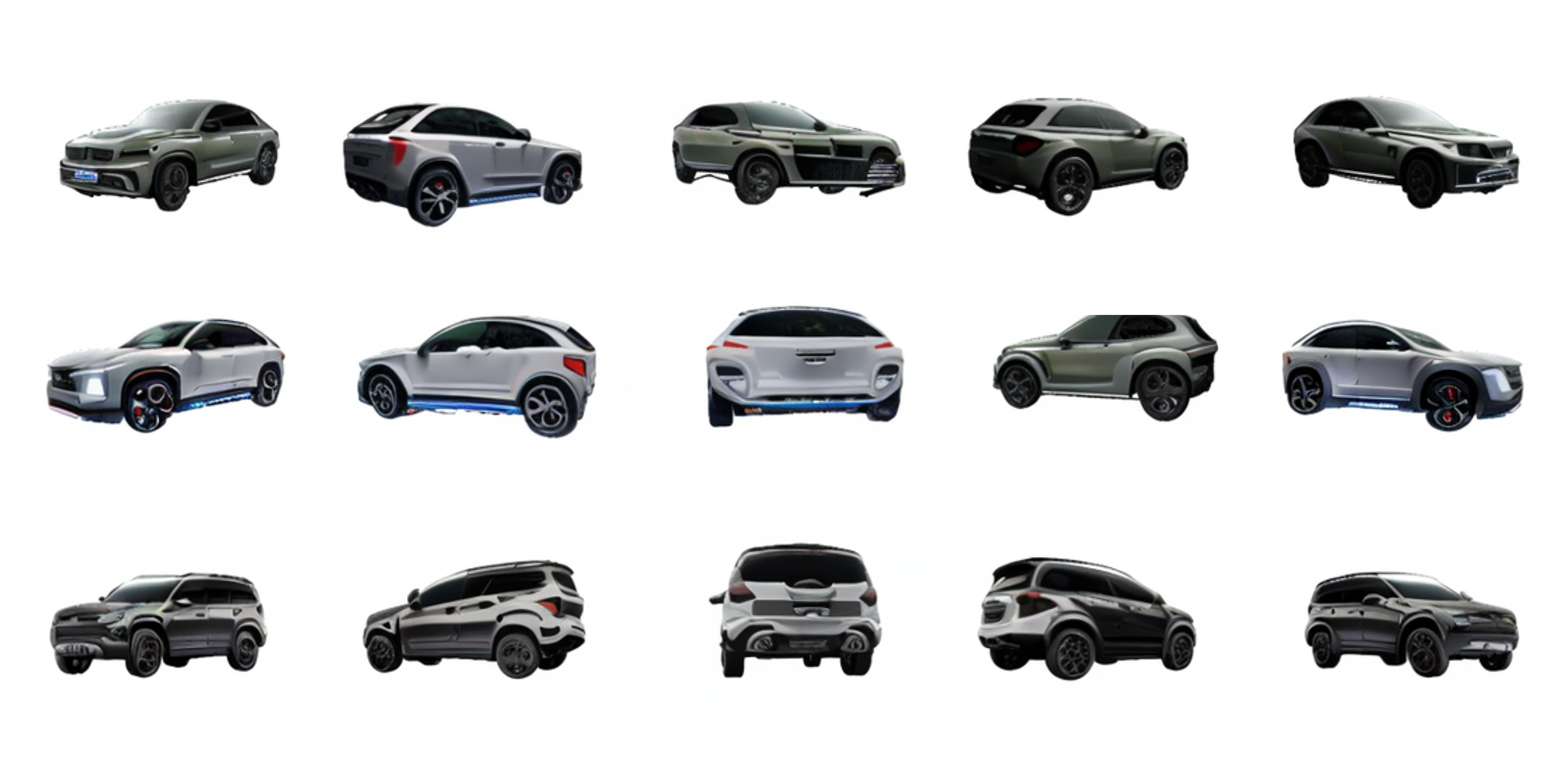} 
\caption{Multi-view images generated from the primary rendering.}
\label{fig:multi_view}
\end{figure}

This Agent is responsible for generating the 3D Point Cloud from the rendering. In the context of our workflow, it takes the high-fidelity primary visual rendering as input and processes it to extract the necessary geometric information. This involves identifying key features and structures within the rendering, which are then mapped into a 3D space to create a detailed Point Cloud representation of the vehicle's body. We use 2 off-the-shelf models, a 3D diffusion model Zero-1to3 \citep{zero1to3} and a 3D reconstruction model VGGT \citep{vggt}, to achieve this. The 3D diffusion model generates a set of standardized multi-view images from the rendering, as shown in Figure \ref{fig:multi_view}. While the 3D reconstruction model integrates and reconstructs them into a unified 3D point cloud capable of accurately representing the vehicle's body surfaces. 

\subsubsection{Aerodynamic Prediction Agent}
With the 3D point cloud generated, the aerodynamic performance validation agent is responsible for predicting the aerodynamic characteristics of the vehicle design. Traditional CFD simulations are accurate but very slow, often taking hours or days for complex designs. To overcome this bottleneck, we employ a data-driven surrogate model in the agent whose core architecture is inspired by the Transolver \citep{Transolver}. Transolver introduces a novel Physics-Attention mechanism, efficiently capturing intrinsic physical correlations from discretized geometries. Specifically, the Transolver architecture decomposes the discretized domain into learnable slices, assigning points with similar physical states to the same slice. These slices are encoded into physics-aware tokens, facilitating effective and computationally efficient attention-based predictions.

\subsubsection{Physics Attention Calculation}
Given a set of point cloud $\mathbf{g}=\{\mathbf{g}_{i}\}_{i=1}^{N}$ with coordinate information of points $N$ and observed quantities $\mathbf{u}$, a linear layer embeds them in deep features $\mathbf{x}=\{\mathbf{x}_{i}\}_{i=1}^{N}$, where each feature of the point cloud contains $C$ channels, i.e. $\mathbf{x}_{i}\in\mathbb{R}^{1\times C}$, and involves both geometry and physics information. To capture physical states under the whole input domain, each point $\mathbf{g}_{i}$ ascribed to $M$ potential slices based on its learned feature $\mathbf{x}_{i}$, which is formalized as follows:
\begin{equation}
	\begin{split}\label{equ:learn_slice}
\{\mathbf{w}_{i}\}_{i=1}^{N} &= \left\{\operatorname{Softmax}\big(\operatorname{Project}\left(\mathbf{x}_{i}\right)\big)\right\}_{i=1}^{N} \\
\mathbf{s}_{j} &= \left\{\mathbf{w}_{i,j}\mathbf{x}_{i}\right\}_{i=1}^{N},
	\end{split}
\end{equation}
where $\operatorname{Project}()$ projects the $C$ channel feature into $M$ weights and yields slice weights $\mathbf{w}_{i}\in\mathbb{R}^{1\times M}$ after the $\operatorname{Softmax}()$ operation. Specifically, $\mathbf{w}_{i,j}$ represents the degree to which the $i$-th point belongs to the $j$-th slice, with $\sum_{j=1}^{M} \mathbf{w}_{i,j}=1$. $\mathbf{s}_{j}\in\mathbb{R}^{N\times C}$ denotes the $j$-th slice feature, which is a weighted combination of the $N$ point features $\mathbf{x}$. Points with similar features will produce similar slice weights and are likely to be assigned to the same slice. To avoid uniform assignment across slices,  $\operatorname{Softmax}()$ are used along the slice dimension to encourage low-entropy and informative slice weight distributions. 

Subsequently, since each slice aggregates points with similar geometric and physical characteristics, through spatially weighted aggregation encode them into physics-aware tokens.
\begin{equation}
	\begin{split}\label{equ:token_encoding}
\mathbf{z}_{j} = \frac{\sum_{i=1}^{N} \mathbf{s}_{j,i}}{\sum_{i=1}^{N}\mathbf{w}_{i,j}} = \frac{\sum_{i=1}^{N} \mathbf{w}_{i,j}\mathbf{x}_{i}}{\sum_{i=1}^{N} \mathbf{w}_{i,j}} ,
	\end{split}
\end{equation}
where $\mathbf{z}_{j}\in\mathbb{R}^{1\times C}$. Each token feature $\mathbf{z}_{j}$ is normalized by dividing the sum of the corresponding slice weights. After encoding from physically consistent slices through spatial aggregation, each token captures the information of a specific physical state.

Formally, the physics attention mechanism among tokens is defined, for a deep feature $\mathbf{x}\in\mathbb{R}^{N\times C}$ embedded from input, firstly decompose it into $M$ physically internal-consistent slices $\mathbf{s}=\{\mathbf{s}_{j}\}_{j=1}^{M}\in\mathbb{R}^{M\times (N\times C)}$ based on learned slice weights $\mathbf{w}\in\mathbb{R}^{N\times M}$. Then, to obtain the specific physics information contained in each slice, aggregate $M$ slices to $M$ physics-aware tokens $\mathbf{z}=\{\mathbf{z}_{j}\}_{j=1}^{M}\in\mathbb{R}^{M\times C}$ by Eq.~\eqref{equ:token_encoding}. The attention mechanism among encoded tokens to capture intricate correlations among different physical states, that is
\begin{equation}
\begin{split}\label{equ:attn}
\mathbf{q}, \mathbf{k}, \mathbf{v} = \operatorname{Linear}(\mathbf{z}), \ \ \mathbf{z}^\prime = \operatorname{Softmax}\left(\frac{\mathbf{q}\mathbf{k}^{\sf T}}{\sqrt{C}}\right)\mathbf{v},
	\end{split}
\end{equation}
where $\mathbf{q}, \mathbf{k}, \mathbf{v}, \mathbf{z}^\prime\in\mathbb{R}^{M\times C}$.
Afterward, transited physical tokens $\mathbf{z}^\prime=\{\mathbf{z}_{j}^\prime\}_{j=1}^{M}$ are transformed back to mesh points by deslicing, which recomposes tokens with slice weights:
\begin{equation}
	\begin{split}\label{equ:deslice}
\mathbf{x}_{i}^\prime & = \sum_{j=1}^{M} \mathbf{w}_{i,j}\mathbf{z}_{j}^\prime,
	\end{split}
\end{equation}
where $1\leq i\leq N$ and each token $\mathbf{z}_{j}^\prime$ is broadcasted to all mesh points during above calculation.

Following the architecture of canonical Transformer, the $l$-th layer can be formalized as:
\begin{equation}
	\begin{split}\label{equ:overall}
\hat{\mathbf{x}}^{l} &= \operatorname{Physics-Attn}\left(\operatorname{LayerNorm}\left({\mathbf{x}}^{l-1}\right)\right) + {\mathbf{x}}^{l-1}\\
{\mathbf{x}}^{l} &= \operatorname{FeedForward}\left(\operatorname{LayerNorm}\left(\hat{\mathbf{x}}^{l}\right)\right) + \hat{\mathbf{x}}^{l},
	\end{split}
\end{equation}
where $l \in \{1,\dots,L\}$, $x^l \in \mathbb{R}^{N \times C}$ is the output of the $l$-th layer, and $x^0 \in \mathbb{R}^{N \times C}$ represents the input deep feature embedded from input geometries $g \in \mathbb{R}^{N \times C_g}$ and initial observation $u \in \mathbb{R}^{N \times C_u}$ via a embedding layer.

\begin{table*}[!tbp]
    \centering
    \setlength{\tabcolsep}{6pt} 
    \renewcommand{\arraystretch}{1.0} 
    \begin{tabular}{l|ccccc}\toprule
    & \multicolumn{5}{c}{\textbf{Aerodynamic Drag Prediction}} \\
    & $\text{MSE} (\times 10^{-2}) \downarrow$ & $\text{MAE} (\times 10^{-1}) \downarrow$ & $R^2 \text{ Score} \uparrow$ & Training Time (h) & Inference Time (s/sample) \\\midrule
    TripNet & 2.6020 & \textbf{4.0300} & 0.9720 & - & - \\
    PointNet & 12.000 & 8.8500 & 0.8260 & \textbf{0.51} & 2.3 \\
    PointNet++ & 7.8130 & 6.7550 & 0.8960 & 0.52 & 2.4 \\
    GCNN & 10.700 & 7.1700 & 0.8740 & 20.71 & \textbf{0.1} \\
    \textbf{Ours} & \textbf{2.1044} & 4.4700 & \textbf{0.9740} & 1.75 & 2.1 \\\midrule\midrule
    & \multicolumn{5}{c}{\textbf{Surface Field Prediction}} \\
    & $\text{MSE} (\times 10^{-2}) \downarrow$ & $\text{MAE} (\times 10^{-1}) \downarrow$ & MAX AE $\downarrow$ & Rel $L_2$ Error (\%) $\downarrow$ & Rel $L_1$ Error (\%) $\downarrow$ \\ \midrule
    TripNet & 4.2300 & 1.1100 & \textbf{5.5200} & 20.3500 & 18.5200 \\
    FIGConvNet & 4.3800 & 1.1300 & 5.7300 & 20.9800 & 18.5900 \\
    RegDGCNN & 9.0100 & 1.5800 & 13.0900 & 28.4900 & 26.2900 \\
    \textbf{Ours} & \textbf{3.9700} & \textbf{0.9600} & 7.9389 & \textbf{14.5600} & \textbf{9.7500} \\\midrule\midrule
    & \multicolumn{5}{c}{\textbf{Velocity Field Prediction}} \\
    & $\text{MSE}\downarrow$ & $\text{MAE}\downarrow$ & MAX AE $\downarrow$ & Rel $L_2$ Error (\%) $\downarrow$ & Rel $L_1$ Error (\%) $\downarrow$ \\\midrule
    $\mathbf{U_{x}}$ & 6.52 & 1.06 & 31.12 & 9.77 & 7.42 \\
    $\mathbf{U_{y}}$ & 2.16 & 0.92 & 24.09 & 32.93 & 27.42 \\
    $\mathbf{U_{z}}$ & 2.10 & 0.97 & 25.88 & 33.43 & 31.45 \\
    $\mathbf{U}$ & 6.27 & 1.85 & 29.32 & 6.63 & 10.78 \\
    \bottomrule
    \end{tabular}
    \caption{Performance comparison of our proposed model against several baseline methods for aerodynamic analysis. The evaluation covers three key prediction tasks: overall drag coefficient $\mathbf{C_d}$, surface pressure field $\mathbf{p}$, and surrounding velocity field $\mathbf{v}$, assessed using various standard metrics.}
    \label{tab:result}
\end{table*}

\subsubsection{Model Architecture} 
\begin{figure}[t]
    \centering
    \includegraphics[width=0.9\columnwidth]{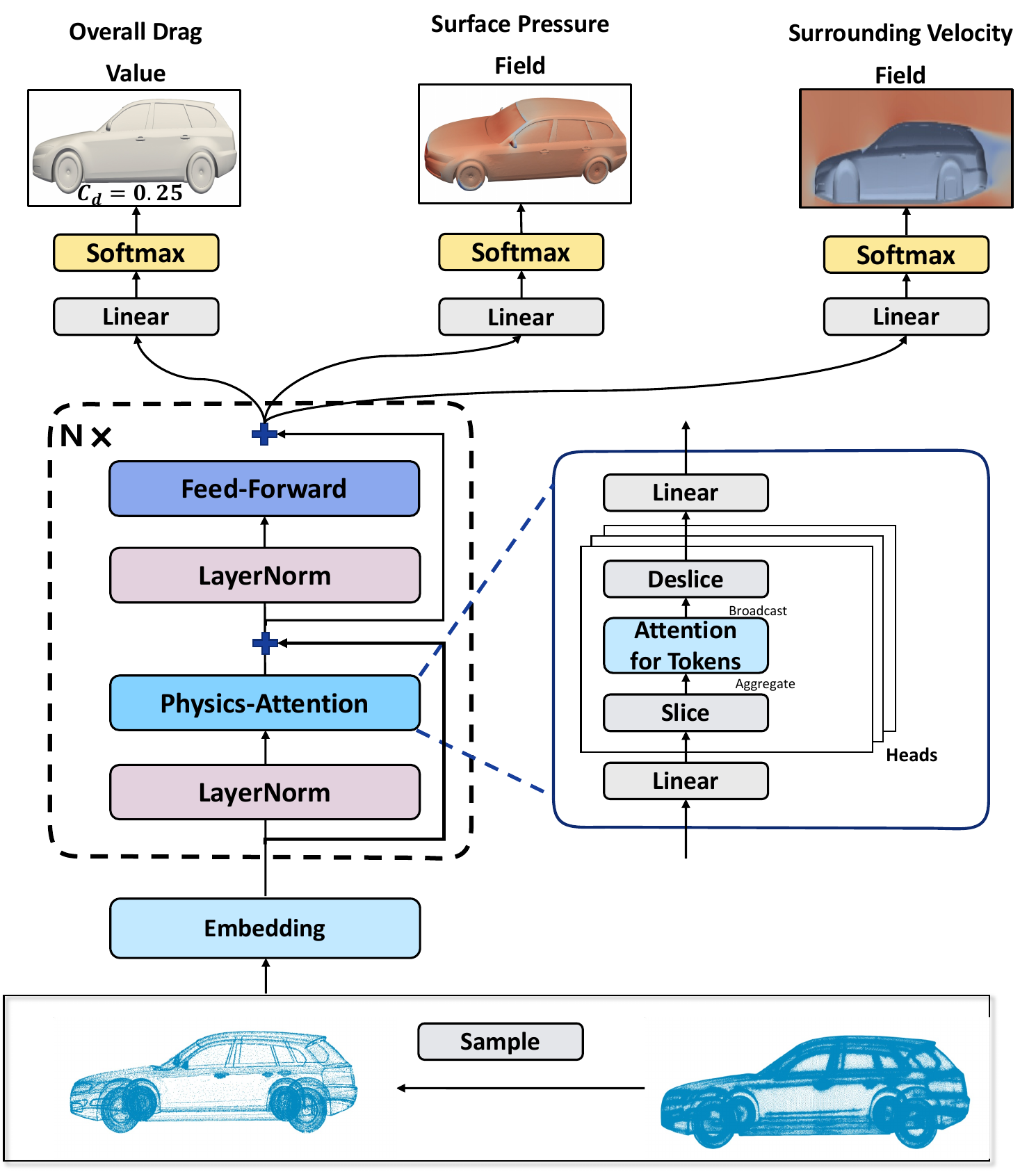} 
    \caption{The model architecture. Sampled points are embedded and processed by $L$ Transolver layers, then fed into three prediction heads for drag, pressure, and velocity field prediction.}
    \label{model}
\end{figure}

As shown in Figure~\ref{model}, to address the computational challenges of large-scale 3D vehicle point clouds, we begin by applying a adaptive sampling strategy based on regional division of the raw input point cloud $\mathbf{g} \in \mathbb{R}^{N \times C_g}$ and its associated physical observations $\mathbf{u} \in \mathbb{R}^{N \times C_u}$. This strategy preserves key curvature features (upper part) and density-driven redundancy removal (lower part), effectively reducing the number of points while preserving essential geometric and physical information, enabling scalable inference.

The sampled points are then processed by an embedding layer that transforms the geometric coordinates and physical quantities into a unified deep feature representation $\mathbf{x}^0 \in \mathbb{R}^{N \times C}$. This embedding incorporates both spatial and physics-based context, forming the initial input to the main architecture.

The core of the model consists of a stack of $L$ Transolver layers. Each layer integrates a Physics-Attention mechanism that dynamically slices and aggregates point features into physics-aware tokens. These tokens capture spatial and physical coherence across the domain and are refined through self-attention to learn complex inter-slice interactions. After attention-based token transformation, the information is projected back onto the point-wise representation via deslicing, enabling rich point-level feature updates. Formally, this process follows Eq.~\eqref{equ:overall}, iterating through $L$ layers to obtain the final feature $\mathbf{x}^L$.

The resulting feature $\mathbf{x}^L$ is then passed to three task-specific prediction heads, each implemented as a lightweight feed-forward decoder to predict the overall drag coefficient $\mathbf{C_d}$, surface pressure field $\mathbf{p}$ and surrounding velocity field $\mathbf{v}$.

This unified multi-task architecture enables the model to efficiently infer critical aerodynamic metrics in a single forward pass, substantially accelerating the design iteration cycle for aerodynamic performance evaluation.

\subsubsection{Loss Function} To simultaneously optimize for the three distinct outputs, the model is trained end-to-end using a composite loss function. For the high-dimensional field predictions, we follow the strategy outlined in the Shape-Net Car benchmark, which combines a loss for the surrounding velocity field ($\mathcal{L}_v$) and a loss for the surface pressure field ($\mathcal{L}_p$). Both are calculated using the Relative L2 error:
\begin{equation}
    \mathcal{L}_{\text{field}} = \frac{ \| \mathbf{y} - \hat{\mathbf{y}} \|_2 }{ \| \mathbf{y} \|_2 },
\end{equation}
where $\mathbf{y}$ is the ground-truth field and $\hat{\mathbf{y}}$ is the model's prediction. For the scalar drag coefficient ($C_d$) prediction, we employ a standard MSE loss, denoted as $\mathcal{L}_{C_d}$. The total loss function, $\mathcal{L}_{\text{total}}$, is therefore defined as:
\begin{equation}
\label{equ:loss}
    \mathcal{L}_{\text{total}} = \lambda_v \mathcal{L}_{v} + \lambda_s \mathcal{L}_{p} + \lambda_{C_d} \mathcal{L}_{C_d},
\end{equation}
By minimizing this composite loss, the model learns a balanced and unified feature representation capable of supporting all three prediction tasks.

\section{Experiments}
\subsection{Experimental Setup}
The core reasoning engine for each agent was powered by "GPT-4o". We trained and tested our Aerodynamic Prediction Agent on the DrivAerNet dataset \citep{drivaernet}, which has 8,000 3D car models, each model comes with detailed CFD simulation data, including pressure, velocity, and drag coefficient. We implemented the model with $L=6$ Transolver layers, an embedding dimension of $C=256$, and $M=64$ physics-aware slices. The model was trained for 200 epochs using the Adam optimizer with a learning rate of 0.001. The loss weights in Eq. \ref{equ:loss} were set to $\lambda_v = 1.0$, $\lambda_p = 1.0$, and $\lambda_{C_d} = 0.1$.

\subsection{Evaluation of Aerodynamic Prediction Performance}
\subsubsection{Main Result} In Table \ref{tab:result}, we present the results of three key prediction tasks: aerodynamic drag, surface field, and velocity field.

For aerodynamic drag prediction, our model is compared against methods including TripNet, PointNet, PointNet++, and GCNN. While TripNet achieves a slightly lower MAE, our model demonstrates highly competitive performance with a low MSE of 2.1044 and high $R^2$ score 0.9740, significantly outperforming PointNet and GCNN on this metric. At the same time, our model achieves a training time of 1.75 hours, which is significantly faster than GCNN's 20.71 hours, while maintaining a near-instantaneous inference time of 2.1 seconds per sample compared with time-consuming CFD simulations.

In the surface field prediction task, our model shows superior performance by outperforming all listed baselines across key metrics. Notably, our model achieves the lowest Relative $L_2$ Error (14.56\%) and Relative $L_1$ Error (9.75\%), which indicates a more accurate prediction of the pressure distribution across the vehicle's body compared to these other methods.

For the velocity field prediction, our model demonstrates strong performance across all three components of the velocity field ($\mathbf{U_x}$, $\mathbf{U_y}$, and $\mathbf{U_z}$). The MSE values are 6.52, 2.16, and 2.10 respectively, with corresponding MAE values of 1.06, 0.92, and 0.97. The overall velocity field $\mathbf{U}$ achieves a MSE of 6.27 and an MAE of 1.85, indicating that our model effectively captures the complex fluid dynamics around the vehicle models.

\subsubsection{Ablation Study}
We conducted an ablation study to investigate the impact of different sampling strategies and sample sizes on the performance of our surrogate model. We compared random sampling (\textit{Rand}), curvature-based sampling (\textit{Curvature}), and our proposed adaptive sampling strategy (\textit{Adaptive}), evaluated at sample sizes of 10k, 20k, and 50k points. As shown in Table~\ref{tab:ablation_study}, Adaptive sampling outperforms both Curvature and Rand across all scales, with performance steadily improving as the sample size increases,demonstrating its efficiency and accuracy.
\begin{table}[h]
    \centering
    \small
    \begin{tabular}{lccc}
        \toprule
        \textbf{Sampling Method} & \textbf{MSE} $\downarrow$ & \textbf{MAE} $\downarrow$ & \textbf{$R^{2}$} $\uparrow$ \\
        \midrule
        Curvature-10k & 1.9258 & 5.0010 & 0.8865 \\
        Curvature-20k & 1.9075 & 4.9100 & 0.8848 \\
        Curvature-50k & 1.8937 & 4.9006 & 0.8906 \\ \hline
        Adaptive-10k  & 1.8333 & 4.8410 & 0.8971 \\
        Adaptive-20k  & 1.7793 & 4.6105 & 0.8974 \\
        Adaptive-50k  & \textbf{1.7673} & \textbf{4.6007} & \textbf{0.9012} \\ \hline
        Rand-10k      & 1.9842 & 5.1070 & 0.8859 \\
        Rand-20k      & 1.8795 & 4.9440 & 0.8890 \\
        Rand-50k      & 1.8203 & 4.7284 & 0.8948 \\
        \bottomrule
    \end{tabular}
    \caption{Ablation study comparing different sampling methods and numbers of sampled points. Higher $R^{2}$ and lower MSE/MAE indicate better performance.}
    \label{tab:ablation_study}
\end{table}

\subsection{Evaluation of Conceptual and Creative Design}
\subsubsection{Human Evaluation}

\begin{figure}[h]
\centering
\includegraphics[width=0.9\columnwidth]{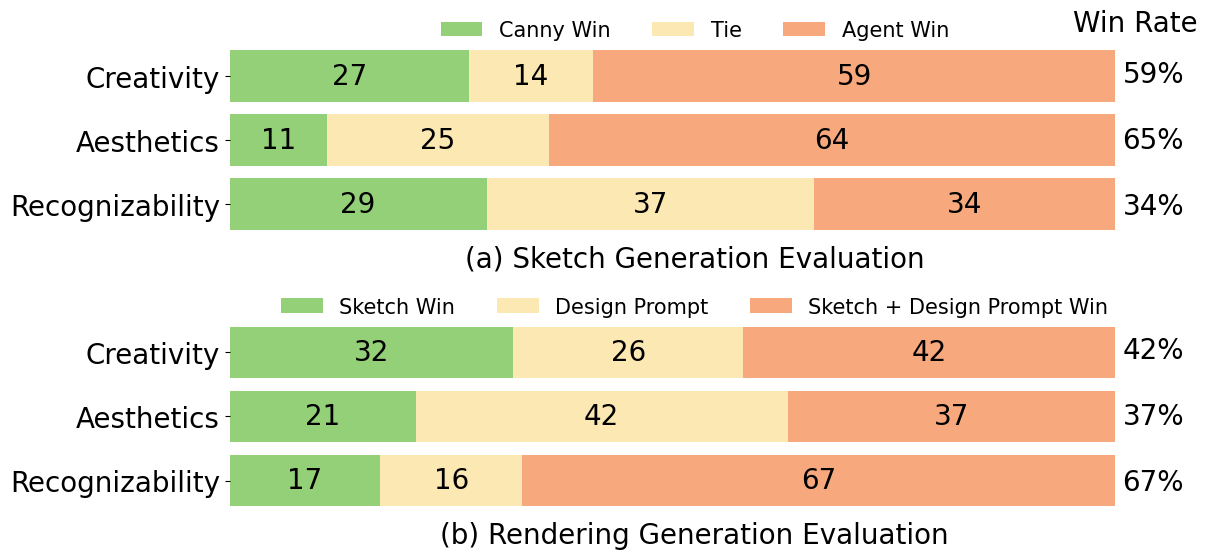}
\caption{Human evaluation results for concept sketch generation and rendering quality across different input conditions.}
\label{human_eval}
\end{figure}

As shown in Figure \ref{human_eval}, to evaluate the quality of the generated concept sketches, we conducted a human evaluation study involving 10 professional automotive designers. Each participant reviewed a set of 100 concept sketches generated by our Rendering Generation Agent, and ranked them according to three criteria: creativity, aesthetic, and recognizability. Sketches produced by the Canny edge detection were used as a baseline. In addition, we conducted a second evaluation to assess the effectiveness of different input conditions in rendering generation. We tested  three configurations: (1) sketches only, (2) design prompts only, (3) sketches + design prompts.

\subsubsection{Objective Metrics}
Table~\ref{tab:objective_sketch} shows our quantitative evaluation results. CLIPasso sketches notably outperform the Canny baseline in both FID and CLIP-Score. For renderings, combining sketches and prompts achieves the best semantic alignment and diversity.

\begin{table}[h]
\centering\small
\begin{tabular}{lccc}
\toprule
\textbf{Sketch Method} & FID$\downarrow$ & CLIP-S$\uparrow$ & LPIPS$\uparrow$ \\
\midrule
Canny Edge & 47.11 & 26.04 & 0.54  \\
\textbf{Ours (CLIPasso)} & \textbf{40.02} & \textbf{29.40} & \textbf{0.58} \\ \midrule \midrule
\textbf{Rendering Input} & FID$\downarrow$ & CLIP-S$\uparrow$ & LPIPS$\uparrow$ \\
\midrule
Sketch Only & 35.7 & 28.13 & 0.44 \\
Prompt Only & \textbf{32.1} & 32.44 & 0.47 \\
\textbf{Sketch + Prompt} & 35.6 & \textbf{39.07} & \textbf{0.52} \\
\bottomrule
\end{tabular}
\caption{Evaluation metrics of concept \textit{sketches} and different input modalities on final \textit{renderings}.}
\label{tab:objective_sketch}
\end{table}

\subsubsection{3D Reconstruction Result}
As shown in table \ref{tab:3d_recon}, our pipeline achieves a 51\% reduction in Chamfer Distance compared with Point-E and more than doubles the margin over the MCC baseline. Meanwhile, the IoU rises from 0.188 to 0.205. These gains confirm that leveraging multi-view diffusion priors before geometry fusion enables sharper edge recovery and fewer ghost points, which are crucial for downstream aerodynamic prediction.
\begin{table}[h]
\centering\small
\begin{tabular}{lcccc}
\toprule
\textbf{} & MCC & SJC-I & Point-E & \textbf{Ours} \\
\midrule
CD $\downarrow$ & 0.0641 & 0.0571 & 0.0551 & \textbf{0.0265} \\
IOU $\uparrow$ & 0.1386 & 0.1880 & 0.1880 & \textbf{0.2047} \\
\bottomrule
\end{tabular}
\caption{Quantitative comparison of 3D reconstruction performance with respect to Chamfer Distance (CD) and Intersection-over-Union (IoU).}
\label{tab:3d_recon}
\end{table}

\section{Conclusion}
In this paper, we presented a novel two-stage multi-agent framework for automotive design, integrating conceptual design and aerodynamic performance validation. The framework orchestrate specialized agents, enabling rapid translation of user requirements into high-fidelity visual concepts and efficient aerodynamic analysis using a lightweight surrogate model. Our experimental results demonstrate the effectiveness of the proposed approach in automating the design process and achieving accurate aerodynamic predictions, significantly enhancing the efficiency of automotive design workflows.

\bibliography{aaai2026}


\end{document}